\documentstyle[12pt]{article}

\def\mib#1{\mbox{\boldmath $#1$}}
\begin{document}
\ \ \ 
\vspace{2.0cm}
\begin{center}
\noindent
{\Large \bf Non-Abelian Gauge Configuration with a Magnetic Field 
Concentrated at a Point\\}
\vspace{1.0cm}

Minoru HIRAYAMA\footnote{hirayama@sci.toyama-u.ac.jp}, 
Takeshi HAMADA$^{\dagger,}$\footnote{hamada@hep.s.kanazawa-u.ac.jp}\\and 
Masafumi HASEGAWA\footnote{hsgw@jodo.sci.toyama-u.ac.jp}\\
\vspace{1.5cm}
{\it Department of Physics, Toyama University, Toyama 930 \\
${}^{\dagger}$Department of Physics, Kanazawa University, Kanazawa  920-11 }
\end{center}
\par
\ \ \ 

\vfill
\begin{center}
{\bf Abstruct}
\end{center}
\vspace{0.5cm}

A specific SU(2) gauge configuration yielding a magnetic field concentrated 
at a point is investigated. Its relation to the Aharonov-Bohm gauge potential 
and its cohomological meaning in a three dimensional space are clarified. 
Quantum mechanics of a spinning particle in such a gauge configuration is 
briefly discussed.

\vfill
\begin{center}
{\small To be published in Progress of Theoretical Physics Vol. 98, 
No. 6 (1997)}\
\ \\
\ \\
\ \\ 
\end{center}

\newpage
In a recent paper,$^{1)}$ two of the present authors and H.-M. Zhang 
considered static $SU(2)$ gauge configurations of the following type:
$$ A_i(\mib x)\equiv \frac{g}{\hbar c}a_i(\mib x)= \frac{\lambda(r)}{r^2}
\epsilon_{ijk}x_j T_k, $$
$$ i=1,2,3,\ \ \ \mib x={}^t(x_1,x_2,x_3),\ \ \  r=|\mib x|,\eqno(1) $$
where $a_i(\mib x), g$ and $T_k$ are the gauge potential, the gauge coupling 
constant and a representation of the generator of $su(2)$ satisfying
$$ [T_i,T_j]=i\epsilon_{ijk}T_k, $$
$$ {\rm tr}(T_iT_j)=\sigma\delta_{ij},\ \ \ \sigma>0,\eqno(2) $$
respectively. Under the Wu-Yang Ansatz, (1), the field equation is given 
by$^{2)}$
$$ r^2\frac{d^2}{dr^2}\lambda(r)=\lambda (r)\{\lambda (r)-1\}
\{\lambda (r)-2 \}. \eqno(3) $$
Obvious solutions of (3) are $\lambda (r)=0,1$ and $2$. The Yang-Mills 
action, $S$, is maximum 
for $\lambda (r)=0$ and $\lambda(r)=2$ with $S=0$ and minimum for 
$\lambda(r)=1$ with $S=-\infty$. 
The case $\lambda(r)=0$ is uninteresting. The case $\lambda(r)=1$ corresponds 
to the much-discussed point-like $SU(2)$ magnetic monopole which is known as 
the Wu-Yang monopole.$^{1)-3)}$ 
In this article, we concentrate on the case
$$ \lambda(r)=2. \eqno(4) $$
In this case, the gauge potential can be rewritten as a pure gauge:$^{1)}$
$$ A_i(\mib x)=iU(\hat{\mbox{\mib x}})\partial_i U^\dagger
(\hat{\mbox{\mib x}}),\eqno(5) $$
$$ U(\hat{\mbox{\mib x}})=e^{i\pi{\hat{\mbox{\mib x}}}\cdot\mib T},
\ \ \ \hat{\mbox{\mib x}}=\frac{\mib x}{r}.\eqno(6) $$
It can be seen that $U(\hat{\mbox{\mib x}})$ satisfies 
$U(\hat{\mbox{\mib x}})=U^{-1}(\hat{\mbox{\mib x}})=
U^\dagger(\hat{\mbox{\mib x}})$. This configuration is gauge-equivalent to the
 trivial configuration $\mib A(\mib x)=0$ in the space 
$\mib R^3\setminus\{\bf 0\}$. It is, however, nontrivial in $\mib R^3$ 
including the origin $\bf 0$. The field strength 
$F_{ij}(\mib x)=\partial_i A_j(\mib x)-\partial_j A_i 
(\mib x)-i[A_i(\mib x),A_j(\mib x)]$ is calculated to be
$$ F_{ij}(\mib x)=-4\delta(r^2)\epsilon_{ijk}T_k.\eqno(7) $$
The singularity of $F_{ij}({\mib x})$ at the origin implies the property 
$\partial_i\partial_jU(\hat{\mbox{\mib x}})|_{\mib x=\bf 0}\ne\partial_j
\partial_iU(\hat{\mbox{\mib x}})|_{\mib x=\bf 0}$,\ $i\ne j$. 
To better understand the meaning of the magnetic field 
$B_i(\mib x)\equiv\frac12\epsilon_{ijk}F_{jk}(\mib x)=-4\delta(r^2)T_i$
 which is nonvanishing only at the origin, we temporalily introduce a small
 scale parameter $s$ and replace the factor $r^{-2}$ in $A_i(\mib x)$ by
 $(r^2+s^2)^{-1}$. Then, the field strength $F_{ij}(\mib x)$ becomes
 $-4s^2(r^2+s^2)^{-2}\epsilon_{ijk}T_k$. Definig a projection of the magnetic
 field by $b_i(\mib x)={\rm tr}\{(\hat{\mbox{\mib x}}\cdot\mib T)
B_i(\mib x)\}/\sigma$, we obtain
$$ \mib b(\mib x)=-{\rm grad}_{\mib x}\int_{\mib R^3} d^3\mib x^{\prime}
 \frac{\rho(r^{\prime})}{|\mib x-\mib x^\prime |},\eqno(8) $$
$$ \rho(r)=\frac{2}{\pi}\frac{s^2(r^2-s^2)}{r(r^2+s^2)^3}.\eqno(9) $$
The function $\rho(r)$ can be interpretted as the density of the projected
 magnetic charge. The $\rho(r)$ is positive in the exterior region $r>s$ and
 negative in the interior region $r<s$. The total positive and negative
 charges in the respective regions are given by $\int_{r>s}\rho(r)d^3{\mib x}
=-\int_{r<s}\rho(r)d^3{\mib x}=1$. In the limit $s\rightarrow0$, the above
 type of magnetic charge distribution gives rise to the magnetic field of the
 prescribed property. Through a similar discussion to the above, we find,
 in contrast with the case of the Wu-Yang monopole,$^{3)}$ the Bianchi identity
$$ \epsilon_{ijk}[D_i,F_{jk}(\mib x)]=0,\ \ \ D_i=\partial_i-iA_i(\mib x),
\eqno(10) $$
is not violated.

 The gauge configuration that we are discussing should be compared with
 Aharonov and Bohm's one.$^{4)}$ The Aharonov-Bohm gauge potential,
 $\mib A^{AB}(\mib x)$, can be expressed by our $\mib A(\mib x)$ in the
 following way. In $\mib R^3$, it is given by
\begin{eqnarray*}
 \mib A^{AB}(\mib x)&=&\frac{\alpha}{x^2_1+x^2_2}\left( \begin{array}{c}
 x_2\\ -x_1 \\0 \end{array} \right) \\
   &=&\frac{\alpha}{4\sigma}\int^{\infty}_{-\infty}\frac{dZ}{|\mib x-\mib Z|}
{\rm tr}\{T_3\mib A(\mib x-\mib Z)\},
\end{eqnarray*} 
$$ \mib x\in \mib R^3,\ \ \ \alpha=\rm const.,\ \ \ \mib Z={}^t(0,0,Z). 
\eqno(11)$$
It is a superposition of $\mib{A(x-Z)}$ which is singular at the point 
$\mib Z$ on the $x_3$-axis. In $\mib R^2$, the $\mib A^{AB}(\mib x)$ is more
 simply given by
\begin{eqnarray*}
 A^{AB}_{i}(\mib x) &=& \frac{\alpha}{x^2_1+x^2_2}\epsilon_{ij}x_j, \\
 &=& \frac{\alpha}{2\sigma}{\rm tr}\{T_3A_i(\mib x)\}|_{x_3=0},
\end{eqnarray*}
$$ \epsilon_{ij}=-\epsilon_{ji},\ \ \ \epsilon_{12}=1,\ \ \ \mib x \in 
\mib R^2,\ \ \ i=1,2.\eqno(12)$$

 We next consider a cohomological meaning of the configuration 
$\mib A(\mib x),\mib x\in \mib R^3$.
Denoting the $p$th de Rham cohomology group of the space $X$ by $H^p(X)$,
 the existence of the Aharanov-Bohm gauge potential is due to the nontrivial
 $H^1(\mib R^2\setminus\{\bf 0\})$. On the other hand, we know that, for the
 space $M\equiv{\mib R}^3\setminus\{\bf 0\}$, $H^1(M)$ is trivial but
 $H^2(M)$ is nontrivial.$^{5)}$ As an example, we investigate the 2-form
 $\omega$ defined by
$$ \omega={\rm tr}\{U(\hat{\mbox{\mib x}})A_i(\mib x)A_j(\mib x)\}dx_i
\wedge dx_j.\eqno(13) $$
It is straightforward to obtain
$$ d\omega=0,\ \ \ {\mib x}\in M. \eqno(14) $$
The period of $\omega$ for a sphere surrounding the origin, however, is
 proportional to the integral of ${\rm tr}\{({\hat{\mbox{\mib x}}}
\cdot{\mib T})U(\hat{\mbox{\mib x}})\}$ on the sphere and does not vanish.
 We see, through de Rham's first theorem,$^{5)}$ that $\omega$ cannot be given
 as an exact form. We conclude that $\omega$ belongs to $H^2(M)$ and that any
 closed 2-form $\lambda$ on $M$ can be written as $\lambda=d\omega_1+a\omega$
 with an appropriate 1-form $\omega_1$ and a constant $a$. We thus see that
 $U(\hat{\mbox{\mib x}})$ and $\mib A(\mib x)$ are convenient quantities to
 describe $H^2(M)$.

 For our $\mib A(\mib x)$, a loop integral $\int_{\gamma}\mib A(\mib x)\cdot
 d\mib x$ for a loop $\gamma \subset M$ is in general nonvanishing. For
 example, for $\gamma_{\theta}\equiv\{(r\sin\theta\cos\varphi,r\sin\theta\sin
\varphi,r\cos\theta)|0\le \varphi < 2\pi,r,\theta:{\rm fixed}\}$, we have 
$\int_{\gamma_{\theta}}\mib A(\mib x)\cdot d\mib x=-4\pi T_3\sin^2\theta$
 which is not equal to a multiple of $2\pi$ in general. We see, however, that
 the loop variable $V(\gamma)$ defined by 
$$ V(\gamma)=Pe^{i\int_{\gamma}\mib A(\mib x)\cdot d\mib x},\ \ \ P:{\rm path
\ ordering}, \eqno(15)$$
is equal to 1:
$$ V(\gamma)=1.\eqno(16)$$
This result is obtained with the help of the non-Abelian Stokes' thorem,
$^{6)-8)}$
$$ V(\gamma)={\cal P}{\rm exp}\left( i\int_{S} d\sigma_{ij} u(\mib x) F_{ij}
(\mib x) u^{\dagger}(\mib x)\right),\eqno(17) $$
where ${\cal P},S,d\sigma_{ij}$ and $u(\mib x)$ are a certain 2-dimensional
 ordering factor, a surface with $\partial S=\gamma$, a surface element of $S$
 and an $\mib x$-dependent unitary matrix, respectively. By (7) and (17), we
 easily understand that $V(\gamma)=1$ if the surface $S$ does not contain the
 origin. When the origin $\bf 0$ lies on $S$, we have $V(\gamma)={\rm exp}
[-4\pi iu({\bf 0})(\mib n\cdot\mib T)u^{\dagger}(\bf 0)]$, where $\mib n$ is
 the normal of $S$ at $\bf 0$. Since all the eigenvalues of $2u({\bf 0})
(\mib n\cdot\mib T)u^{\dagger}(\bf 0)$ are integral, we conclude 
$V(\gamma)=1$ in this case, too.

 One more interesting property of $\mib A(\mib x)$ is that it yields an
 angular momentum satisfying a desired algebra. If we define $\mib j$ by
\begin{eqnarray*}
 \mib j &=& U(\hat{\mbox{\mib x}})\left( \mib x\times\frac1i \nabla \right)
 U^{\dagger}(\hat{\mbox{\mib x}}) \\
        &=& \mib x\times\frac1i(\nabla-i\mib A) 
\end{eqnarray*}
$$\ \ \ \ \ \ \ = \frac1i \mib x\times\nabla-\frac{2}{r^2} \mib x\times
(\mib x\times \mib T),\eqno(18)$$ 
it satisfies the relation
 $[j_{l},j_{m}]=i\epsilon_{lmn}\{j_{n}-4x_{n}({\mib x}\cdot{\mib T})\delta
 (r^2)\}$. Noticing the relation $r^2\delta(r^2)=0$, we have
$$ [j_{l},j_{m}]=i\epsilon_{lmn}j_{n}\eqno(19) $$ 
even at the origin.

 In the above, we have considered a gauge configuration yielding a magnetic
 field at one point. 
Of course, any configration of the form (5) with $U$ singular at a point will
 exhibit similar properties to (7) and (19). If we consider $\mib A^{\nu}
(\mib x)=iU^{\nu}(\hat{\mbox{\mib x}})\{\nabla U^{\nu\dagger}(\hat{\mbox
{\mib x}})\}$ with $U^{\nu}(\hat{\mbox{\mib x}})=e^{i\nu \hat{\mbox{\mib x}}
\cdot\mib T},0\le \nu <2\pi$, we are led to $\mib A^{\nu}(\mib x)=[(\hat
{\mbox{\mib x}} \times \mib T)(1-\cos\nu)-\{\hat{\mbox{\mib x}} \times (\hat
{\mbox{\mib x}} \times \mib T)\}\sin\nu]r^{-1}$. In contrast with our $\mib A
(\mib x)$, the $\mib A^{\nu}(\mib x),\ \nu \ne 0,1,$ does not satisfy the
 Wu-Yang Ansatz and is not a solution of the field equation. If we do not
 restrict ourselves to the solutions of the Yang-Mills field equation, we can
 think of many interesting configurations. For instance, it is clear that the
 $A_i^{(N)}(\mib x)$ defined by
$$ A_i^{(N)}(\mib x)=iU^{(N)}\partial_i U^{(N)\dagger},\eqno(20) $$
$$ U^{(N)}=U_1U_2\cdots U_{N},\ \ U_n={\rm exp}\left\{i\pi\frac{\mib x-
\mib r_n}{|\mib x-\mib r_n|}\cdot\mib T\right\},\eqno(21) $$
causes a magnetic field located at the points $\mib r_1,\mib r_2,\cdots,
\mib r_N$. An appropriate limit of the gauge potential of this type yields the
 Aharonov-Bohm potential as is seen in (11).

 We now turn to a brief discussion of quantum mechanics of a spinning particle
 put in the gauge configuration given by (1) and (4). The Hamiltonian of the
 system is
$$ {\cal H}=-\frac{\hbar^2}{2m}\{\mib\sigma\cdot(\nabla-i\mib A(\mib x))\}
^2+V(r)$$ 
$$ \ \ \ \ \ \ \ \ \ \ \ =-\frac{\hbar^2}{2m}\{(\nabla-i\mib A)^2+\mib\sigma
\cdot\mib B(\mib x)\}+V(r),\eqno(22)$$
where $m$ and $\mib\sigma=(\sigma_1,\sigma_2,\sigma_3)$ denote the mass of the
 particle and the Pauli matrices, respectively. The potential term $V(r)$ is
 assumed to be independent of the spin $\frac12 \hbar \mib \sigma$ and the
 isospin $\mib T$ and dependent only on $r$. The nontriviality of this example
 is manifest in the term $\mib\sigma\cdot\mib B(\mib x)=-4(\mib\sigma\cdot
\mib T)\delta(r^2)$.
Since the $\cal H$ can be written as
$$ {\cal H}=U(\hat{\mbox{\mib x}}){\cal H}_0 U^{\dagger}(\hat{\mbox{\mib x}}),
 $$ 
$$ {\cal H}_0=-\frac{\hbar^2}{2m}(\mib\sigma\cdot\nabla)^2+V(r),\eqno(23) $$
its eigenfunction $\psi(\mib x)$ takes the form
$$ \psi(\mib x)=U(\hat{\mbox{\mib x}})\varphi(\mib x)v.\eqno(24) $$
In (24), $\varphi(\mib x)$ is an eigenfunction of $-(\hbar^2/2m)\nabla^2+V(r)$
 and is regular at the origin, $v$ describing the spin and the isospin degrees
 of freedom. Note that $\varphi(\mib x)v$ satisfies the relation $(\mib\sigma
\cdot\nabla)^2\{\varphi(\mib x)v\}=\{\nabla^2\varphi(\mib x)\}v$. The angular
 momentum $\hbar\mib J$ of the system is given by
$$ \hbar\mib J=\hbar\left(\mib j+\frac{\mib\sigma}2 \right)\eqno(25) $$
with $\mib j$ defined by (18). The commutativity of $\mib J$ with $\cal H$ is
 assured not on $\varphi(\mib x)v$ but on $\psi(\mib x)$:
$$ [{\cal H},\mib J]\psi(\mib x)=0.\eqno(26) $$
If we respect this property, we are forced to make use of $\psi(\mib x)$ which
 is not single-valued unless $\varphi(\mib x)$ vanishes at the origin. 
>From (19) and (25), the components of $\hbar\mib J$ satisfy the desired
 algebra of angular momentum. Because of the afore-mentioned property
 $\partial_i\partial_jU(\hat{\mbox{\mib x}})|_{\mib x=\bf 0}\ne\partial_j
\partial_iU(\hat{\mbox{\mib x}})|_{\mib x=\bf 0}$, $i\ne j$, we see that the
 canonical commutation relation $[p_i,p_j]=0,\ p_i=(\hbar/i)\partial_i$, is
 violated unless $\varphi(\mib x)$ vanishes at the origin:
$$[p_i,p_j]\psi(\mib x)|_{\mib x=\bf 0}\ne 0\ \ (i \ne j)\ \ {\rm if}\ \varphi
({\bf 0})\ne 0. \eqno(27)$$ 
On the other hand, if the $\varphi(\mib x)$ with $\varphi(\bf 0)\ne 0$ is
 excluded, the completeness of the set $\{\psi(\mib x)\}$ and the
 self-adjointness of $\cal H$ will be lost.

 In a future communication, we will discuss whether the multi-valuedness of
 $\psi(\mib x)$ and the violation of the canonical commutation relation at the
 origin cause physical effects or not.

\vspace{0.5cm}
\begin{center}
{\large\bf References}\\
\end{center}
\noindent
1) M. Hirayama, H.-M.Zhang and T. Hamada, Prog. Theor. Phys., {\bf 97}, (1997),
 679.\\
2) A. Actor, Rev. Mod. Phys. {\bf 51} (1979), 461.\\
3) T. T. Wu and C. N. Yang, in {\it Properties of Matter under Unusual
 Conditions}, ed. H.

$\!\!$Mark and S. Feshbach (Interscience, New York, 1968).\\
4) Y. Aharonov and D. Bohm, Phys. Rev. {\bf 115}, 1959, 485.\\
5) H. Flanders, {\it Differential Forms with Applications to the Physical
 Sciences}(Academic

$\!\!$Press, New York, 1963).\\ 
6) N. E. Brali\'c, Rev. {\bf D22}, 1980, 3090.\\
7) Y. Aref'eva, Theor. Math. Phys. {\bf 43} (1980), 353.\\ 
8) B. Broda, in {\it Advanced Electromagnetism : Foundation, Theory and
 Applications}, ed. T.

$\!\!$Barrett and D. Grimes (World Sci. Pub. Co., Singapore,1995).\\

\end{document}